\documentclass[aps,prl,showpacs,twocolumn,lineno,groupedaddress]{revtex4}  
\usepackage{graphicx}  
\usepackage{dcolumn}   
\usepackage{bm}        
\usepackage{amssymb}   

\def\met{\mbox{${\hbox{$E$\kern-0.6em\lower-.1ex\hbox{/}}}_T$}} 

\begin{document}


\hspace{5.2in} \mbox{Fermilab-Pub-05-046-E}

\title{Measurement of the $p\overline{p} \rightarrow W\gamma + X$ cross section
at $\sqrt{s} = 1.96$ TeV and $WW\gamma$ anomalous coupling limits}
%
\author{                                                                      
V.M.~Abazov,$^{35}$                                                           
B.~Abbott,$^{72}$                                                             
M.~Abolins,$^{63}$                                                            
B.S.~Acharya,$^{29}$                                                          
M.~Adams,$^{50}$                                                              
T.~Adams,$^{48}$                                                              
M.~Agelou,$^{18}$                                                             
J.-L.~Agram,$^{19}$                                                           
S.H.~Ahn,$^{31}$                                                              
M.~Ahsan,$^{57}$
% ...
\\                                                                            
\vskip 0.30cm                                                                 
\centerline{(D\O\ Collaboration)}                                             
\vskip 0.30cm                                                                 
}                                                                             
\address{                                                                     
\centerline{$^{1}$Universidad de Buenos Aires, Buenos Aires, Argentina}       
\centerline{$^{2}$LAFEX, Centro Brasileiro de Pesquisas F{\'\i}sicas,         
                  Rio de Janeiro, Brazil}                                     
\centerline{$^{3}$Universidade do Estado do Rio de Janeiro,                   
                  Rio de Janeiro, Brazil}                                     
\centerline{$^{4}$Instituto de F\'{\i}sica Te\'orica, Universidade            
                  Estadual Paulista, S\~ao Paulo, Brazil}                     
\centerline{$^{5}$University of Alberta, Edmonton, Alberta, Canada,           
               Simon Fraser University, Burnaby, British Columbia, Canada,}   
\centerline{York University, Toronto, Ontario, Canada, and                    
         McGill University, Montreal, Quebec, Canada}                         
\centerline{$^{6}$Institute of High Energy Physics, Beijing,                  
                  People's Republic of China}                                 
\centerline{$^{7}$University of Science and Technology of China, Hefei,       
                  People's Republic of China}                                 
\centerline{$^{8}$Universidad de los Andes, Bogot\'{a}, Colombia}             
\centerline{$^{9}$Center for Particle Physics, Charles University,            
                  Prague, Czech Republic}                                     
\centerline{$^{10}$Czech Technical University, Prague, Czech Republic}        
\centerline{$^{11}$Institute of Physics, Academy of Sciences, Center          
                  for Particle Physics, Prague, Czech Republic}               
\centerline{$^{12}$Universidad San Francisco de Quito, Quito, Ecuador}        
\centerline{$^{13}$Laboratoire de Physique Corpusculaire, IN2P3-CNRS,         
                 Universit\'e Blaise Pascal, Clermont-Ferrand, France}        
\centerline{$^{14}$Laboratoire de Physique Subatomique et de Cosmologie,      
                  IN2P3-CNRS, Universite de Grenoble 1, Grenoble, France}     
\centerline{$^{15}$CPPM, IN2P3-CNRS, Universit\'e de la M\'editerran\'ee,     
                  Marseille, France}                                          
\centerline{$^{16}$Laboratoire de l'Acc\'el\'erateur Lin\'eaire,              
                  IN2P3-CNRS, Orsay, France}                                  
\centerline{$^{17}$LPNHE, IN2P3-CNRS, Universit\'es Paris VI and VII,         
                  Paris, France}                                              
\centerline{$^{18}$DAPNIA/Service de Physique des Particules, CEA, Saclay,    
                  France}                                                     
\centerline{$^{19}$IReS, IN2P3-CNRS, Universit\'e Louis Pasteur, Strasbourg,  
                France, and Universit\'e de Haute Alsace, Mulhouse, France}   
\centerline{$^{20}$Institut de Physique Nucl\'eaire de Lyon, IN2P3-CNRS,      
                   Universit\'e Claude Bernard, Villeurbanne, France}         
\centerline{$^{21}$III. Physikalisches Institut A, RWTH Aachen,               
                   Aachen, Germany}                                           
\centerline{$^{22}$Physikalisches Institut, Universit{\"a}t Bonn,             
                  Bonn, Germany}                                              
\centerline{$^{23}$Physikalisches Institut, Universit{\"a}t Freiburg,         
                  Freiburg, Germany}                                          
\centerline{$^{24}$Institut f{\"u}r Physik, Universit{\"a}t Mainz,            
                  Mainz, Germany}                                             
\centerline{$^{25}$Ludwig-Maximilians-Universit{\"a}t M{\"u}nchen,            
                   M{\"u}nchen, Germany}                                      
\centerline{$^{26}$Fachbereich Physik, University of Wuppertal,               
                   Wuppertal, Germany}                                        
\centerline{$^{27}$Panjab University, Chandigarh, India}                      
\centerline{$^{28}$Delhi University, Delhi, India}                            
\centerline{$^{29}$Tata Institute of Fundamental Research, Mumbai, India}     
\centerline{$^{30}$University College Dublin, Dublin, Ireland}                
\centerline{$^{31}$Korea Detector Laboratory, Korea University,               
                   Seoul, Korea}                                              
\centerline{$^{32}$CINVESTAV, Mexico City, Mexico}                            
\centerline{$^{33}$FOM-Institute NIKHEF and University of                     
                  Amsterdam/NIKHEF, Amsterdam, The Netherlands}               
\centerline{$^{34}$Radboud University Nijmegen/NIKHEF, Nijmegen, The          
                  Netherlands}                                                
\centerline{$^{35}$Joint Institute for Nuclear Research, Dubna, Russia}       
\centerline{$^{36}$Institute for Theoretical and Experimental Physics,        
                  Moscow, Russia}                                             
\centerline{$^{37}$Moscow State University, Moscow, Russia}                   
\centerline{$^{38}$Institute for High Energy Physics, Protvino, Russia}       
\centerline{$^{39}$Petersburg Nuclear Physics Institute,                      
                   St. Petersburg, Russia}                                    
\centerline{$^{40}$Lund University, Lund, Sweden, Royal Institute of          
                   Technology and Stockholm University, Stockholm,            
                   Sweden, and}                                               
\centerline{Uppsala University, Uppsala, Sweden}                              
\centerline{$^{41}$Lancaster University, Lancaster, United Kingdom}           
\centerline{$^{42}$Imperial College, London, United Kingdom}                  
\centerline{$^{43}$University of Manchester, Manchester, United Kingdom}      
\centerline{$^{44}$University of Arizona, Tucson, Arizona 85721, USA}         
\centerline{$^{45}$Lawrence Berkeley National Laboratory and University of    
                  California, Berkeley, California 94720, USA}                
\centerline{$^{46}$California State University, Fresno, California 93740, USA}
\centerline{$^{47}$University of California, Riverside, California 92521, USA}
\centerline{$^{48}$Florida State University, Tallahassee, Florida 32306, USA} 
\centerline{$^{49}$Fermi National Accelerator Laboratory, Batavia,            
                   Illinois 60510, USA}                                       
\centerline{$^{50}$University of Illinois at Chicago, Chicago,                
                   Illinois 60607, USA}                                       
\centerline{$^{51}$Northern Illinois University, DeKalb, Illinois 60115, USA} 
\centerline{$^{52}$Northwestern University, Evanston, Illinois 60208, USA}    
\centerline{$^{53}$Indiana University, Bloomington, Indiana 47405, USA}       
\centerline{$^{54}$University of Notre Dame, Notre Dame, Indiana 46556, USA}  
\centerline{$^{55}$Iowa State University, Ames, Iowa 50011, USA}              
\centerline{$^{56}$University of Kansas, Lawrence, Kansas 66045, USA}         
\centerline{$^{57}$Kansas State University, Manhattan, Kansas 66506, USA}     
\centerline{$^{58}$Louisiana Tech University, Ruston, Louisiana 71272, USA}   
\centerline{$^{59}$University of Maryland, College Park, Maryland 20742, USA} 
\centerline{$^{60}$Boston University, Boston, Massachusetts 02215, USA}       
\centerline{$^{61}$Northeastern University, Boston, Massachusetts 02115, USA} 
\centerline{$^{62}$University of Michigan, Ann Arbor, Michigan 48109, USA}    
\centerline{$^{63}$Michigan State University, East Lansing, Michigan 48824,   
                   USA}                                                       
\centerline{$^{64}$University of Mississippi, University, Mississippi 38677,  
                   USA}                                                       
\centerline{$^{65}$University of Nebraska, Lincoln, Nebraska 68588, USA}      
\centerline{$^{66}$Princeton University, Princeton, New Jersey 08544, USA}    
\centerline{$^{67}$Columbia University, New York, New York 10027, USA}        
\centerline{$^{68}$University of Rochester, Rochester, New York 14627, USA}   
\centerline{$^{69}$State University of New York, Stony Brook,                 
                   New York 11794, USA}                                       
\centerline{$^{70}$Brookhaven National Laboratory, Upton, New York 11973, USA}
\centerline{$^{71}$Langston University, Langston, Oklahoma 73050, USA}        
\centerline{$^{72}$University of Oklahoma, Norman, Oklahoma 73019, USA}       
\centerline{$^{73}$Brown University, Providence, Rhode Island 02912, USA}     
\centerline{$^{74}$University of Texas, Arlington, Texas 76019, USA}          
\centerline{$^{75}$Southern Methodist University, Dallas, Texas 75275, USA}   
\centerline{$^{76}$Rice University, Houston, Texas 77005, USA}                
\centerline{$^{77}$University of Virginia, Charlottesville, Virginia 22901,   
                   USA}                                                       
\centerline{$^{78}$University of Washington, Seattle, Washington 98195, USA}  
}                                                                             
\date{\today}

\begin{abstract}
The $WW\gamma$ triple gauge boson coupling parameters are studied using
$p\overline{p} \rightarrow \ell\nu\gamma + X (\ell = e,\mu)$ events at
$\sqrt{s} = 1.96$ TeV.  The data were collected with the D\O\ detector 
from an integrated
luminosity of 162 pb$^{-1}$ delivered by the Fermilab Tevatron
Collider.  The cross section times branching fraction for
$p\overline{p} \rightarrow W(\gamma) + X \rightarrow \ell\nu\gamma + X$
with $E_T^\gamma > 8$ GeV and 
$\Delta {\cal R}_{\ell\gamma} >$ 0.7 is 
$14.8 \pm 1.6 $(stat)$ \pm 1.0 $(syst) $\pm 1.0 $(lum) pb.
The one-dimensional
95\% confidence level limits on anomalous couplings are
$-0.88 < \Delta\kappa_\gamma < 0.96$ and $-0.20 < \lambda_\gamma < 0.20$.
\end{abstract}

\maketitle

The $W\gamma$ final states observed at hadron colliders provide an 
opportunity to study the self-interaction of electroweak bosons at the
$WW\gamma$ vertex.  
The standard model (SM) description of electroweak physics is based on 
SU(2)$_L\otimes$U(1)$_Y$ gauge symmetry and specifies the 
$WW\gamma$ coupling.  
In the SM, production of a photon in association with a $W$ boson
occurs due to radiation of a photon from an incoming quark, from the $W$ boson
due to direct $WW\gamma$ coupling, or from the outgoing $W$ boson decay 
lepton.
To allow
for non-SM couplings, a CP-conserving 
effective Lagrangian can be written with two coupling parameters:  $\kappa_\gamma$ and
$\lambda_\gamma$ \cite{Hagi,BaurMC}.
The SM predicts
$\Delta\kappa_\gamma \equiv \kappa_\gamma - 1 = 0$ and $\lambda_\gamma = 0$.  
Non-standard couplings cause the effective Lagrangian to violate partial wave 
unitarity at high energies; it is necessary to introduce a form-factor
with scale $\Lambda$ for each of the coupling parameters.  The form-factors
are introduced 
via the ansatz
$\lambda \rightarrow \lambda / (1 + \hat{s}/\Lambda^2)^2$
with $\sqrt{\hat{s}}$ the $W\gamma$ invariant mass.  
In this analysis, the scale $\Lambda$ is set to 2 TeV.  For sufficiently
small values of $\Lambda$ the dependence on $\Lambda$ is relatively small.
Deviations from the SM $WW\gamma$ couplings would cause an increase in 
the total $W\gamma$ production cross section and would enhance
the production of photons with high transverse energy.  

Limits on the $WW\gamma$ coupling parameters have been
previously reported by the D\O\ \cite{RunIDZero} and CDF 
\cite{RunICDF} collaborations using 
direct observation of $W\gamma$ final states
in data collected from hadron collisions at the 
Fermilab Tevatron collider and by the UA2 \cite{UA2} collaboration using the 
$Sp\overline{p}S$ collider at CERN.  
Searches for $W^+W^-$ final states at D\O\ \cite{RunIDZeroWW} and 
CDF \cite{RunICDFWW} have also been used
to test $WW\gamma$ and $WWZ$ coupling parameters simultaneously.
Similarly, experiments at the CERN LEP collider constrain the $WW\gamma$
and $WWZ$ coupling parameters simultaneously through observations of
$W^+W^-$, single-$W$ boson, and single-$\gamma$ final states 
in electron-positron collisions \cite{LEP}.
Observation of
$b \rightarrow s\gamma$ decays by the CLEO collaboration 
has also been used to constrain
the coupling parameters \cite{Cleo}.  

The analyses discussed here use the D\O\ detector to observe 
$p\overline{p} \rightarrow \ell\nu\gamma + X (\ell = e $ or $\mu)$ events in 
collisions at $\sqrt{s} = 1.96$ TeV at the Fermilab Tevatron collider.
The data samples used for the electron and muon channels correspond to
integrated luminosities of 162 pb$^{-1}$ and 134 pb$^{-1}$, respectively.
The D\O\ detector \cite{d0det} features an inner tracker
surrounded by a liquid-argon/uranium calorimeter and a muon spectrometer.
The inner tracker consists of a silicon microstrip tracker (SMT) and 
a central fiber
tracker (CFT), both located within a 2 T superconducting solenoidal magnet.  
The
CFT covers $|\eta| \lesssim 1.8$ and the SMT covers $|\eta| \lesssim 3.0$
\cite{coordsys}.
%
The calorimeter is longitudinally segmented into electromagnetic and
hadronic layers and is housed in three cryostats: a central section
covering $|\eta| \lesssim 1.1$ and two end-cap cryostats that extend
coverage to $|\eta| \lesssim 4.0$.  The muon detectors reside outside the calorimeter
and consist of tracking detectors, scintillation counters, and a 1.8 T
toroidal magnet.  The muon detectors cover to $|\eta| \lesssim 2.0$.  Luminosity is
measured using scintillator arrays located in front of the end-cap cryostats
and covering $2.7 \lesssim |\eta| \lesssim 4.4$.

Candidate events with electron decays of the $W$ boson
($W \rightarrow e\nu$) are collected
using a suite of single electron triggers that require electromagnetic
clusters in the calorimeter with at least 11 GeV of transverse energy ($E_T$).  
Offline electron identification requires the candidate electrons
to be in the central calorimeter ($|\eta| < 1.1$), isolated in the calorimeter, 
have shower profiles consistent
with those of electromagnetic objects, and have a track found in the tracking
detectors 
matched to the calorimeter
cluster.  Similarly, photons are identified as central electromagnetic 
calorimeter
clusters without a matched track that are isolated both in the calorimeter
and in the tracking detectors.
To suppress events 
with final state radiation of the photon from the outgoing lepton, and to
avoid collinear singularities in calculations, the
photon is required to be separated from the electron in $\eta-\phi$ space 
($\Delta {\cal R} = \sqrt{(\eta_\gamma - \eta_{\ell})^2 + 
                          (\phi_\gamma - \phi_{\ell})^2} > 0.7$).  
Events used in this analysis are required to have $E_T^e > 25$ GeV, 
$E_T^\gamma > 8$ GeV, missing transverse energy using the full calorimeter
\met $> 25$ GeV, and
$M_T > 40$ GeV$/c^2$, where $M_T$ is the transverse mass 
$\sqrt{2E_T^e \met (1-\cos{\phi^{e\nu}})}$ of the electron and $\met$
vectors which are separated by $\phi^{e\nu}$ in azimuth.

Candidate events with muon decays of the $W$ boson ($W \rightarrow \mu\nu$) are collected 
using a suite of single muon triggers that require a high $p_T$ track in 
the muon
detectors and a high $p_T$ track in the central tracking
detectors.  Offline muon identification additionally 
restricts muon candidates to the
full central tracking acceptance ($|\eta| < 1.6$), requires matched central tracks, and 
imposes timing cuts to reduce
backgrounds from cosmic and beam halo muons.  Events with more than one
identified muon are rejected to reduce backgrounds from 
$Z \rightarrow \mu\mu(\gamma)$.  
Events
are required to have $p_T^\mu > 20$ GeV$/c$, 
$E_T^\gamma > 8$ GeV, \met $> 20$ GeV, and there is no $M_T$ requirement for
this analysis.
Photon identification is the same for both electron and muon analyses.

The dominant background for both decay channels is $W+$jet production where
a jet mimics a photon.  The contribution of this background is estimated by
using a large multijet data sample to measure the probability of jets
to mimic photons.
Some fraction of multijet events contains true photons, and this fraction
has previously been seen to increase with increasing
transverse energy as $1 - e^{a-bE_T}$ \cite{directPho}.
The systematic uncertainty on the probability of a jet
being misidentified as a photon is 
taken to be the full difference between ignoring the presence of
true photons in the multijet data sample and estimating their contribution
with the above functional form.  
The method described above
is dependent on agreement between the jet energy calibration and 
the electromagnetic energy calibration; as a check of the accuracy of the
jet energy calibration, the method is repeated using jet-like objects that have a high
fraction of calorimeter energy in the electromagnetic layers.  This
yields a background estimate consistent with the method based on jets.

A second class of background events comes from processes which produce
an electron or muon, an electron that is misidentified as a photon, and 
missing transverse energy.
This background, labeled ${\ell}eX$, is small for the muon channel
since very few processes produce a high $E_T$ muon and an electron.
However, this background is significant for the electron channel
since $Z (\rightarrow ee)+$jet (with a mismeasured jet leading to apparent
missing transverse energy) processes have a relatively large cross section.
To reduce this background, an additional criterion
 on the invariant mass
of the electron and photon candidates is imposed, and events with 
$70 < M_{e\gamma} < 110$ GeV$/c^2$ are rejected.  In both the electron and muon
analyses, the ${\ell}eX$ background is estimated by reversing the track
match requirement on the photon candidate ({\em i.e.} require a matched track)
in $W\gamma$ candidate events.
The number of ${\ell}eX$ events in which
the electron is isolated and does not have a matched track 
(and therefore is misidentified as a photon) is
then estimated using the known track matching and track isolation 
inefficiencies.

Small backgrounds from $Z\gamma$, where one lepton from the $Z$ decay is not
reconstructed, and $W \rightarrow \tau\nu\gamma$, where the $\tau$ decays
into an electron or muon, are 
estimated from Monte Carlo samples.  The background estimates and numbers of
events observed in the data are summarized in Table \ref{tab:numEvents}.

\begin{table}
\caption{\label{tab:numEvents}Summary of estimated backgrounds and numbers of events selected in 
each channel.}
\begin{ruledtabular}
\begin{tabular}{cr@{$\,\pm\,$}lr@{$\,\pm\,$}l}
 & \multicolumn{2}{c}{$e\nu\gamma$ Channel} & 
   \multicolumn{2}{c}{$\mu\nu\gamma$ Channel} \\
\hline \\
Luminosity & \multicolumn{2}{c}{162 pb$^{-1}$} & 
             \multicolumn{2}{c}{134 pb$^{-1}$} \\
$W +$ jet background events & 58.7 & 4.5 & 61.8 & 5.1 \\
$\ell e X$ background events& 1.7 &  0.5 & 0.7 & 0.2 \\
$W\gamma \rightarrow \tau\nu\gamma$ background events& 0.42 & 0.02 & 1.9 & 0.2 \\
$Z\gamma \rightarrow \ell\ell\gamma$ background events & \multicolumn{2}{c}{-} & 6.9 & 0.7 \\
\hline
Total background events& 60.8 & 4.5 & 71.3 & 5.2 \\
Selected events & \multicolumn{2}{c}{112} & \multicolumn{2}{c}{161} \\
Total signal events& 51.2 & 11.5 & 89.7 & 13.7 \\
\end{tabular}
\end{ruledtabular}
\end{table}

The efficiencies of the triggers and the lepton identification cuts are 
measured using $Z \rightarrow ee, \mu\mu$ events.  Efficiencies for
electrons are $0.96 \pm 0.02$ for the trigger,  $0.84 \pm 0.01$ for 
the calorimeter identification requirements, and $0.78 \pm 0.01$ for the
track match requirement.  
For muons, the trigger efficiency is $0.74 \pm 0.01$, the offline
reconstruction efficiency is $0.77 \pm 0.02$, and the efficiency of the track
match requirement is $0.98 \pm 0.01$.  
The efficiency of the requirement of no more than one muon in muon candidate events
is estimated to
be $0.942 \pm 0.004$ by counting the fraction of $Z \rightarrow ee$ events
containing a muon.  
The track isolation efficiency used for the ${\ell}eX$ background estimation
is measured using $Z \rightarrow ee$ events and is $0.95 \pm 0.01$.
The efficiency of the calorimeter 
requirements in photon identification
is estimated using a full {\sc geant3} simulation of the 
detector \cite{GEANT}.  
The probability for unrelated tracks to 
overlap with the photon and cause it to fail the track isolation requirements
is measured using $Z \rightarrow ee$ events by measuring
the probability of an electron to have nearby tracks after the event
is rotated in $\phi$ by ninety degrees.  The overall efficiency for photon
identification is $0.81 \pm 0.01$.  The total efficiencies
are $0.51 \pm 0.02$ for the electron channel and $0.43 \pm 0.01$ for 
the muon channel.

The acceptances due to the kinematic and geometric requirements in the
analyses are calculated
using a Monte Carlo generator \cite{BaurMC} that fully models $W\gamma$
production to leading order in quantum chromodynamics (QCD) and
electroweak couplings and allows anomalous coupling values to be 
set.  The detector response is simulated using a 
parameterized detector simulation.
The effects of higher order QCD processes are accounted for by the
introduction of a $K$-factor of $1.335$ \cite{BaurMC}, and the transverse
momentum spectrum of the $W$ boson is simulated using parton showers in
{\sc pythia} \cite{PYTHIA}.  The detector acceptance calculation has 
a very small dependence on the simulation of the transverse momentum of the
$W$ boson.
The CTEQ6L parton distribution function (PDF) \cite{CTEQ6} 
is used for the proton and anti-proton. 
The acceptances are $0.045 \pm 0.002$ for the electron 
channel and $0.102 \pm 0.003$ for the muon channel with the uncertainties
dominated by the PDF uncertainty.

The measured cross sections times branching fractions
$\sigma(p\overline{p} \rightarrow W(\gamma) + X \rightarrow \ell\nu\gamma + X)$
with $E_T^\gamma>8$ GeV and $\Delta {\cal R}_{\ell\gamma}>0.7$ are 
$13.9 \pm 2.9 $(stat)$ \pm 1.6 $(syst)$ \pm 0.9 $(lum) pb for the electron 
channel and 
$15.2 \pm 2.0 $(stat)$ \pm 1.1 $(syst)$ \pm 1.0 $(lum) pb
for the muon channel.  
The three components  of the cross section uncertainty are:  statistics; systematic effects associated with the background subtraction, acceptance calculation, and
object identification; and the systematic uncertainties in the luminosity 
measurement.
Combining events from the two decay channels and accounting for 
correlations in the systematic uncertainties yields a combined cross
section times branching fraction of 
$14.8 \pm 1.6 $(stat)$ \pm 1.0 $(syst)$ \pm 1.0 $(lum)~pb.
The SM prediction calculated by the Monte Carlo generator using the 
$K$-factor and the CTEQ6L PDF is $16.0 \pm 0.4$~pb, 
where the uncertainty is due to PDF uncertainty.  The prediction is in 
agreement with the measurements.

\begin{figure}
\includegraphics[scale=0.47]{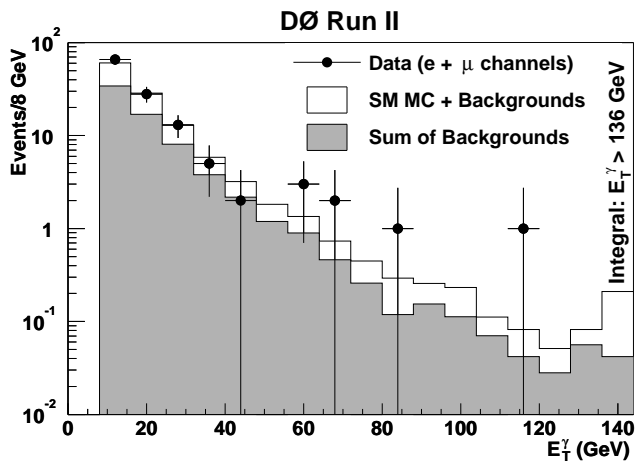}
\caption{
\label{fig:phoET}
The photon $E_T$ spectrum for the $W\gamma$ candidates with 
$M_T(W,\gamma) > 90$ GeV$/c^2$.  The points
with error bars are the data.  The open histogram is the sum of the SM 
Monte Carlo prediction and the background estimate.  The background estimate 
is shown as the shaded histogram.  The right-most bin shows the sum of all
events with photon $E_T$ above 136 GeV.
}
\end{figure}

\begin{figure}
\includegraphics[scale=0.47]{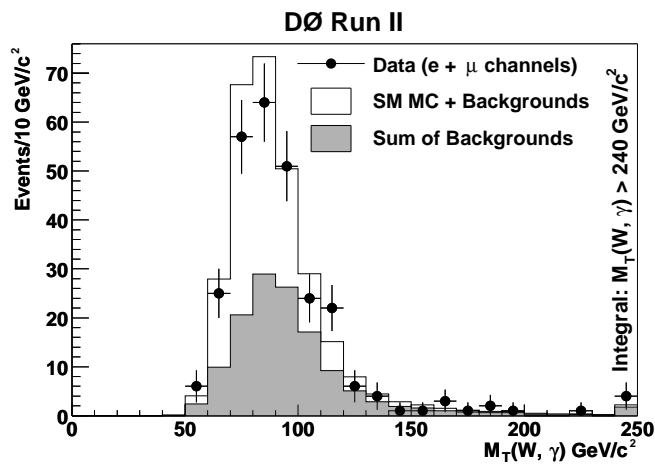}
\caption{
\label{fig:ThreeMass}
The $M_T(W, \gamma)$ spectrum for the $W\gamma$ candidates.  The points
with error bars are the data.  The open histogram is the sum of the SM 
Monte Carlo prediction and the background estimate.  The background estimate 
is shown as the shaded histogram.  The right-most bin shows the sum of
all events with $M_T(W, \gamma)$ above 240 GeV.
}
\end{figure}

The photon $E_T$ spectrum of the candidate events is shown with
the background estimation and the SM expectation in Fig.~\ref{fig:phoET}.
The distribution is described well by the SM, and no
enhancement of the photon $E_T$ spectrum is seen at high transverse energy.
Limits on anomalous couplings are determined by performing a binned
likelihood fit to the photon $E_T$ spectrum.
The effect of anomalous couplings is more pronounced at high $W\gamma$
transverse mass, $M_T(W, \gamma)$, so only
events with $M_T(W, \gamma) > 90$ GeV$/c^2$ are used for the distributions
in the likelihood fit.  The $M_T(W, \gamma)$ distribution before this requirement  
is shown in Fig.~\ref{fig:ThreeMass}. 
  Monte Carlo distributions of the photon $E_T$
spectrum are generated with
a range of anomalous coupling values, and the likelihood of the data 
distribution being consistent with the generated distribution is calculated.  The
uncertainties in the background estimates, efficiencies, acceptances, and
the luminosity are included in the likelihood calculation using Gaussian
distributions.

\begin{figure}
\includegraphics[scale=0.47]{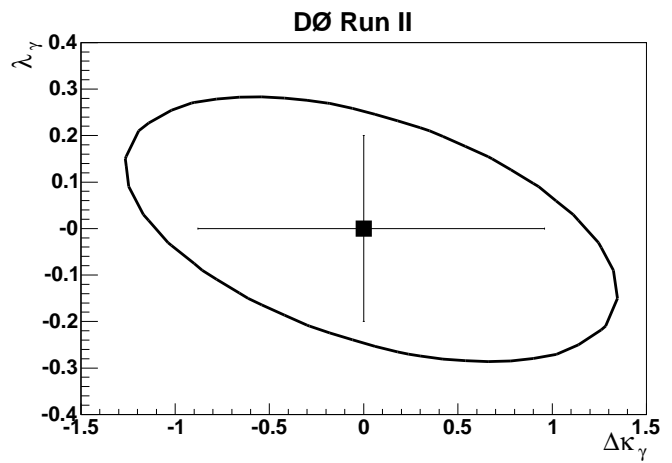}
\caption{
\label{fig:cont}
Limits on the $WW\gamma$ coupling parameters 
$\Delta\kappa_\gamma$ and $\lambda_\gamma$.
The point indicates the SM value with the error bars showing the 95\%
CL intervals in one dimension.  The ellipse represents the 
two-dimensional 95\% CL exclusion contour.
}
\end{figure}

The limits on the $WW\gamma$ coupling parameters are shown in 
Fig.~\ref{fig:cont}, with  
the contour showing
the two-dimensional 95\% confidence level (CL) exclusion limits for the 
coupling parameters, the point representing the Standard Model value and
the error bars showing the one-dimensional 95\%
CL intervals.  The one-dimensional exclusion
limits on each parameter are $-0.88 < \Delta\kappa_\gamma < 0.96$ and 
$-0.20 < \lambda_\gamma < 0.20$, where the limit on $\Delta\kappa_\gamma$ 
assumes $\lambda_\gamma$ is fixed to the SM value and vice versa and
$\Lambda = 2$~TeV.  

In summary, the cross section times branching fraction for the process
$p\overline{p} \rightarrow W(\gamma) + X \rightarrow \ell\nu\gamma + X$ 
with $E_T^\gamma > 8$~GeV and $\Delta {\cal R}_{\ell\gamma} >$ 0.7 is measured
to be $14.8 \pm 1.6 $(stat)$ \pm 1.0 $(sys)$ \pm 1.0 $(lum)~pb using
the D\O\ detector during Run II
of the Tevatron.  The measured cross section is in agreement with the
SM expectation of $16.0 \pm 0.4$~pb.  Limits at the
95\% confidence level on anomalous
$WW\gamma$ couplings are extracted using the photon transverse energy
spectrum and are $-0.88 < \Delta\kappa_\gamma < 0.96$ and
$-0.20 < \lambda_\gamma < 0.20$.
These limits represent the most stringent constraints on anomalous $WW\gamma$
couplings obtained by direct observation of $W\gamma$ production.

%
We thank the staffs at Fermilab and collaborating institutions, 
and acknowledge support from the 
DOE and NSF (USA),
CEA and CNRS/IN2P3 (France),
Ministry of Education and Science, Agency for Atomic 
   Energy and RF President Grants Program (Russia),
CAPES, CNPq, FAPERJ, FAPESP and FUNDUNESP (Brazil),
Departments of Atomic Energy and Science and Technology (India),
Colciencias (Colombia),
CONACyT (Mexico),
KRF (Korea),
CONICET and UBACyT (Argentina),
The Foundation for Fundamental Research on Matter (The Netherlands),
PPARC (United Kingdom),
Ministry of Education (Czech Republic),
Canada Research Chairs Program, CFI,
Natural Sciences and Engineering Research Council and 
WestGrid Project (Canada),
BMBF and DFG (Germany),
Science Foundation Ireland,
A.P.~Sloan Foundation,
Research Corporation,
Texas Advanced Research Program,
Alexander von Humboldt Foundation,
and the Marie Curie Fellowships.
%


\begin {thebibliography}{999}
\bibitem{coordsys} The D\O\ coordinate system is cylindrical with the $z$-axis
along the beamline and the polar and azimuthal angles denoted as $\theta$
and $\phi$ respectively.
%
The pseudorapidity is defined as $\eta = -\ln{\tan{(\theta/2)}}$.
\end{thebibliography}
\end{document}